\documentclass{JHEP3}
\usepackage{MnSymbol}
\title{Mixing with descendant fields in perturbed \\
minimal CFT models}

\author{Armen Poghosyan \\Yerevan Physics Institute\\
Alikhanian Br. 2, 0036 Yerevan, Armenia
\\ E-mail:\email{armenpoghos@yerphi.am}}

\author{Hayk Poghosyan\\Yerevan State University\\
Alex Manoogian 1, 0025 Yerevan, Armenia\\
E-mail:\email{haykpoghos@gmail.com}}

\abstract{
We extend the analysis of the RG trajectory connecting
successive minimal CFT models ${\cal M}_p$ and ${\cal M}_{p-1}$
for $p\gg 1$, performed by A. Zamolodchikov, to the fields
$\varphi_{n,n\pm 3}$. This required a close
investigation of mixing with the descendant fields at the level $2$.
In particular we identify those specific linear combinations of
UV fields which flow to the IR fields $\varphi_{n+3,n}$ and
$\varphi_{n-3,n}$. We report also the results of the calculation
of the same mixing coefficients through the recent
RG domain wall approach by D. Gaiotto. These results are in
complete agreement with the leading order perturbation theory.}

\begin{document}
\section*{Introduction}
In his work \cite{Zamolodchikov:1987} A.~Zamolodchikov
has  investigated two dimensional QFTs (denoted as
${\cal M}_{p,p-1}$) which are perturbations of the
minimal models \cite{BPZ:1984,FQS:1984} ${\cal M}_p$
by the relevant primary
field $\varphi_{1,3}$. In large $p$ limit he has shown
that this theory corresponds to the RG trajectory
connecting two successive minimal models ${\cal M}_p$
and ${\cal M}_{p-1}$. In other words ${\cal M}_{p,p-1}$
interpolates between the theories ${\cal M}_p$
(in UV limit) and ${\cal M}_{p-1}$
(in IR limit)\footnote{Another remarkable aspect of
the $\varphi_{1,3}$ perturbation, namely its
integrability \cite{Zamolodchikov:1987IP}, will not be discussed in
this paper.}.
Zamolodchikov has investigated in details the
renormalization of the fields $\varphi_{n,n}$,
$\varphi_{n,n\pm 1}$, $\varphi_{n,n\pm 2}$
and computed their matrices
of anomalous dimensions. Next to leading order
calculations of the same matrices of anomalous
dimensions have been carried out in the recent paper
\cite{Poghossian:2ndorder}. Note also that the
$N=1$ super-symmetric analogue of this RG flow
is analysed in \cite{Poghossian:1988}.

In this paper we extend Zamolodchikov's analysis to the fields
$\varphi_{n,n\pm 3}$. The relative complexity of our case
is due to strong mixing with the second level descendants
of $\varphi_{n,n\pm 1}$.
The general formula of Zamolodchikov for the matrix of anomalous
dimensions is designed for the case of primary fields (and for the derivatives thereof),
 that is why we will adjust his formula to make it applicable also for more generic
descendant fields. Effectively we deal with the mixing of 10
different fields, hence we need to calculate an $10 \times 10$
matrix of anomalous dimensions. We find that under RG
the UV fields $\varphi_{n,n\pm 3}$ flow to the definite combinations
of the IR fields $\varphi_{n\pm 3,n}$ and the second level
descendants of $\varphi_{n\pm 1,n}$ and explicitly  calculate the
mixing coefficients.
Some times ago a new method of constructing the UV-IR map has been developed \cite{Fre_Que:2005branes,Bru_Rog:2008orb}. The idea is to find a specific conformal
interface (called RG domain wall) between UV and IR theories, which encodes this map.
Recently Gaiotto suggested an algebraic construction for the RG domain wall responsible
for the  ${\cal M}_p \rightarrow {\cal M}_{p-1}$ flow \cite{Gaiotto:2012RGDW} of our current interest.
In this paper we report also the result of calculation
of already mentioned mixing coefficients obtained through the recent
RG domain wall approach by Gaiotto,
which in large $p$ limit completely agrees with our leading order perturbation theory result.
It is worth noting that in contrast to Zamolodchikov's method, the domain wall approach
works perfectly well also in the case of degenerated conformal dimensions, which is
a common feature of the descendant fields.

The paper is organized as follows.

In section \ref{adjustment} we briefly recall Zamolodchikov's
method \cite{Zamolodchikov:1987} of leading order calculation of
the anomalous dimensions in perturbed conformal theories.
We describe here how to adjust his method for the cases with
descendant fields.

Section \ref{3point} is devoted to the calculation of the three
point functions of up to second level descendant fields which
we use in the next section for the calculation of the effective
structure and normalization constants.

In section \ref{anomdim} we present our calculation of the
$10 \times 10 $ matrix of anomalous dimensions.

In section \ref{Comparison} we report the results of
the calculation of mixing coefficients obtained through the
RG domain wall approach suggested by Gaiotto
\cite{Gaiotto:2012RGDW} and get a complete agreement
with the leading order perturbation theory results.

In Appendix we collect few relevant facts about minimal
models of the two dimensional CFT.
\section{Zamolodchikov's theory and its adjustment}
\label{adjustment}
We will briefly present here the leading order perturbation
theory developed by A.~Zamolodchikov to investigate the
renormalization of fields in a conformal field theory perturbed by relevant operators. Denote the action density by
${\cal H}(x,g^i)$, where $g^i$, $i=1,2,\ldots ,n $ are the
(renormalized) coupling constants. It is assumed that
$g^i=0$ corresponds to a CFT and the primary spinless
fields $\Phi_i\equiv \partial {\cal H}/\partial g^i$
are the perturbing operators which are
conventionally normalized as
\begin{eqnarray}
\label{norm}
\left.\langle\Phi_{i}(x)\Phi_{j}(0)\rangle
\right|_{x^2=1}=\delta_{i,j}+O(g^2)
\end{eqnarray}
In the case when the dimensions $\Delta_i$ of these fields satisfy
the conditions $0<\epsilon_i\equiv 1-\Delta_i \ll 1$ and
$g\lesssim \epsilon$, A.~Zamolodchikov has derived a simple expression
for the matrix of the anomalous dimensions
\begin{eqnarray}
\label{Z gamma}
\gamma_i^j(g)=\Delta_i\delta_i^j+C_{ik}^j g^k+O(g^2),
\end{eqnarray}
where $C_{ik}^j$ are related to the structure constant
$C_{ijk}\equiv \langle \Phi_i(1)\Phi_j(0)
\Phi_k(\infty)\rangle |_{g=0}$
\begin{eqnarray}
C_{ik}^j=\pi C_{ijk}
\label{CCrelation}
\end{eqnarray}
Suppose we have a set of primary fields with
close to each other dimensions. Suppose further that
no other field (primary or descendant) of the dimension
approximately equal to those of the set can be
generated by means of the OPE of the fields from our
set with the perturbing fields. Then for the matrix of
anomalous dimensions the same formula
(\ref{Z gamma}) can be used, with the indices $i$ and $j$
running over all the fields of our set. Unfortunately such closed
with respect to OPE sets of primary fields are rare. In most
cases the closed in above mentioned sense sets will
include besides primaries also descendant fields.

The key ingredient of Zamolodchikov's derivation (\ref{Z gamma}) is
the evaluation of the first order perturbative integral
\begin{eqnarray}
\label{Z I}
&&\int d^2y\langle\varphi_i(x,\bar{x})\varphi_j(0,0)
\varphi_k(y,\bar{y})\rangle=
C_{ijk}(x^2)^{\Delta_k-\Delta_i-\Delta_j}\nonumber\\
&&\times \int d^2y(x-y)^{\Delta_j
-\Delta_i-\Delta_k}y^{\Delta_i-
\Delta_j-\Delta_k}
(\bar{x}-\bar{y})^{\Delta_j-\Delta_i
-\Delta_k}\bar{y}^{\Delta_i-\Delta_j
-\Delta_k}\nonumber\\
&&=(x^2)^{1-\Delta_i-\Delta_j-\Delta_k}C_{ijk}I_{ij}^k
\label{pertint}
\end{eqnarray}
where in first equality the standard expression for
the three point functions of primary fields is used and
\begin{eqnarray}
\label{Iijk}
I_{ij}^{k}=\pi\, \frac
{\Gamma(\Delta_i-\Delta_j-\Delta_k+1)\Gamma(\Delta_j-\Delta_i
-\Delta_k+1)\Gamma(2\Delta_k-1)}
{\Gamma(\Delta_i+\Delta_k
-\Delta_j)\Gamma(\Delta_j+\Delta_k-\Delta_i)\Gamma(2-2\Delta_k)}
\end{eqnarray}

It is important to note that the form of $x$ dependence of the
integral (\ref{pertint}) is completely fixed from the
dimensional analysis and remains the same also in the case
when $\varphi_i$ or $\varphi_j$ (or both) are descendants. Only the
numerical coefficient $C_{ijk}I_{ij}^k$ in the general case would be different.
We find it convenient also in general case to factor out the same quantity
$I_{ij}^k$ (\ref{Iijk}) and denote the remaining part as $\tilde{C}_{ijk}$.
$\tilde{C}_{ijk}$ is in some sense an "effective" structure constant.
It can be shown that in large p limit it coincides with the structure constant
of OPE of the respective fields.
 This observation immediately suggests the following strategy
for the calculation of the matrix of anomalous dimensions
in the case when descendant fields are present:
\begin{itemize}
\item{ normalize all the (possibly descendant) fields
as in eq. (\ref{norm})}
\item{calculate the integrals of the three point
functions}
\item{separate factors $I_{ij}^k$ (\ref{Iijk}) from the
overall numerical coefficients and denote the remaining parts
as $\tilde{C}_{ijk}$}
\end{itemize}
then the matrix of the anomalous dimensions would be
\begin{eqnarray}
\label{My gamma}
\gamma_i^j(g)=\Delta_i\delta_i^j+\pi\tilde{C}_{ijk} g^k+O(g^2),
\end{eqnarray}
Let us come back to the main case of our interest, namely to
the theory with a single coupling constant denoted as ${\cal M}_{p,p-1}$.
We are interested in the behavior of the spinless fields $\Phi_{n,m}(x,g)$ ($g$ is the coupling
constant) in ${\cal M}_{p,p-1}$. At $g=0$, $\Phi_{n,m}(x,0)=\varphi_{n,m}(x)$ by
definition ($\varphi_{n,m}(x)$ are the primaries of  ${\cal M}_p$).
Let $n-m=l>0$, $n,m\ll p$. The structure of OPE
with the perturbing field $\varphi_{1,3}$ schematically is
\begin{eqnarray}
\label{1st OPE}
\varphi_{n,m}\varphi_{1,3}=
[\varphi_{n,m}]+
[\varphi_{n,m+2}]+
[\varphi_{n,m-2}]
\end{eqnarray}
where the square brackets stand for the corresponding
conformal family.
The cases when $l=0,1,2$ are analysed in \cite{Zamolodchikov:1987}.
As it is easy to see from (\ref{1st OPE}), the field $\varphi_{n,n}$
by itself alone constitutes a closed in already discussed sense set.
Hence, it does not mix any other field and the matrix of anomalous
dimensions is one dimensional. When $l=1$, the closed set consists
of two primaries $\{\varphi_{n,n+1},\varphi_{n,n-1}\}$. Already
for $l=2$ we have mixing with a descendant (in this case with
a derivative of a primary) and the set is
$\{\varphi_{n,n+2},\partial{\bar\partial}\varphi_{n,n},
\varphi_{n,n-2}\}$\cite{Zamolodchikov:1987}. The next case which is the main
subject of this paper includes $10$ different fields.
Besides the primaries $\varphi_{n,n\pm 3}$, also
second level descendants of the fields $\varphi_{n,n\pm 1}$,
altogether $8$ descendants should be included to get a closed
set.

The RG trajectory of ${\cal M}_{p,p-1}$
has two fixed points at $g=0$ and at $g=g_*$
\cite{Zamolodchikov:1987}
\begin{eqnarray}
\label{2nd point}
2\pi g_* =\sqrt{3}\epsilon+O(\epsilon^2)
\end{eqnarray}
where $\epsilon=2/(p+1)$ is a small parameter.
The fixed points at $g=0$ and $g=g_*$ are described
by the minimal models ${\cal M}_p$ and ${\cal M}_{p-1}$ respectively.
The fields $\Phi_{\alpha}(x,g_*)$ ($\alpha$ is an
index numbering the fields of a closed set) should be
identified with some combination of fields from ${\cal M}_{p-1}$.
To specify this map it is necessary to calculate the
respective matrix of anomalous dimensions. The eigenvalues
of this matrix at $g=g_*$ are the dimensions of those
fields from ${\cal M}_{p-1}$ in term of which the fields
$\Phi_{\alpha}(x,g_*)$ can be expanded. On the other
hand the components of an eigenvector show,
which combination of the fields $\Phi_{\alpha}(x,g_*)$ gives us
the specific field from the IR theory ${\cal M}_{p-1}$ whose
dimension is equal to the respective eigenvalue.
\section{Calculation of three point functions}
\label{3point}
Let us begin with the investigation of the fields $\Phi_{n,n\pm 3}(x,g)$
which at vanishing coupling constant coincide with the
primaries $\varphi_{n,n\pm 3}$ of the theory ${\cal M}_p$. As we
discussed at the end of the previous section we should include
into consideration also the second level descendants of the
fields $\varphi_{n,n\pm 1}$
\begin{eqnarray}
\label{My des}
\mathcal{L}_a\bar{\mathcal{L}}_b\varphi_{n,n\pm1},\qquad\qquad a,b \in \{1,2\}
\end{eqnarray}
where,
\begin{eqnarray}
\label{My L}
&&\mathcal{L}_1\equiv{L_{-1}^2 }\nonumber\\
&&\mathcal{L}_2\equiv{L_{-1}^2  -\frac{2(2\Delta+1)}{3}L_{-2}}
\end{eqnarray}
Here $L_n$   are Virasoro  generators and $\Delta$ is the dimension of  the primary field on which this generators act.
The form of the operator $\mathcal{L}_2$ is chosen so that
the fields $\mathcal{L}_2\varphi$ are quasi primaries.
For the resulting properly normalized $10$ fields
we make the following assignment
\begin{eqnarray}
\label{My f}
&&\Phi_1(x,0)\equiv N_1^{-\frac{1}{2}}\varphi_{n,n+3}(x),
\quad\quad\quad\,
\Phi_2(x,0) \equiv N_2^{-\frac{1}{2}}\mathcal{L}_1
\bar{\mathcal{L}}_1 \varphi_{n,n+1}(x),\nonumber\\
&&\Phi_3(x,0) \equiv N_3^{-\frac{1}{2}}\mathcal{L}_1
\bar{\mathcal{L}}_2 \varphi_{n,n+1}(x),\quad
\Phi_4(x,0)\equiv  N_4^{-\frac{1}{2}}\mathcal{L}_2
\bar{\mathcal{L}}_1 \varphi_{n,n+1}(x),\nonumber\\
&&\Phi_5(x,0) \equiv N_5^{-\frac{1}{2}}\mathcal{L}_2
\bar{\mathcal{L}}_2 \varphi_{n,n+1}(x), \quad
\Phi_6(x,0) \equiv N_6^{-\frac{1}{2}}\mathcal{L}_1
\bar{\mathcal{L}}_1 \varphi_{n,n-1}(x),\nonumber\\
&&\Phi_7(x,0) \equiv N_7^{-\frac{1}{2}}\mathcal{L}_1
\bar{\mathcal{L}}_2 \varphi_{n,n-1}(x),\quad
\Phi_8(x,0) \equiv N_8^{-\frac{1}{2}}\mathcal{L}_2
\bar{\mathcal{L}}_1 \varphi_{n,n-1}(x), \nonumber\\
&&\Phi_9(x,0) \equiv N_9^{-\frac{1}{2}}\mathcal{L}_2
\bar{\mathcal{L}}_2 \varphi_{n,n-1}(x),\quad
\Phi_{10}(x,0)\equiv N_{10}^{-\frac{1}{2}}
\varphi_{n,n-3}(x)
\end{eqnarray}
where the constants $N_1$,\dots,$N_{10}$ are determined
from the normalization condition.
As seen from (\ref{My f}) our fields are either primaries, quasi
primaries or their derivatives.
As we saw in section \ref{adjustment}, to calculate the
matrix of anomalous dimensions we need all three
point functions of the form $\langle \Phi_i
\Phi_j \Phi \rangle|_{g=0}$.
It is sufficient to calculate three point correlation functions
of the primary and quasi primary fields, since we can
always take out the derivatives from the
correlation functions. The basic correlation functions
that should be calculated are of the form
$\langle L_{-2}\varphi_1\varphi_2
\varphi_3\rangle$ and $\langle L_{-2}\varphi_1
L_{-2}\varphi_2\varphi_3\rangle$
where $\varphi_1,\varphi_2,\varphi_3$ are primary fields.
All three point functions of our interest can be easily
derived from the correlators of this kind simply taking
derivatives and using the holomorphic anti-holomorphic
factorization property. Our problem boils down to the
calculation of the correlation functions
$\langle T\varphi_1\varphi_2\varphi_3\rangle$ and
$\langle TT\varphi_1\varphi_2\varphi_3\rangle$ where T is the Energy-Momentum tensor. Such correlators can be computed
using conformal Ward identities \cite{BPZ:1984}. The
results are
\begin{eqnarray}
\label{Tfff}
\langle T(\xi)\varphi_1\varphi_2\varphi_3\rangle=\sum_{i=1}^3
(\frac{\Delta_i}{(\xi-x_i)^2} + \frac{1}{\xi-x_i})\frac{\partial}
{\partial x_i}\langle\varphi_1(x_1)\varphi_2(x_2)
\varphi_3(x_3)\rangle
\end{eqnarray}
and
\begin{eqnarray}
\label{TTfff}
&&\langle T(\xi_1)T(\xi_2)\varphi_1\varphi_2\varphi_3\rangle=
\frac{c}{2(\xi_{1}-\xi_2)^4}\langle\varphi_1(x_1)
\varphi_2(x_2)\varphi_3(x_3)\rangle\\
&&+\left(\frac{2}{(\xi_1-\xi_2)^2}+\frac{1}{\xi_1-\xi_2}\frac{\partial}{\partial \xi_2}+\sum_{i=1}^3(\frac{\Delta_i}{(\xi_1-x_i)^2}
+ \frac{1}{\xi_1-x_i}\frac{\partial}{\partial x_i})\right)
\langle T(\xi_2)\varphi_1(x_1)\varphi_2(x_2)
\varphi_3(x_3)\rangle  \nonumber
\end{eqnarray}
where $c$ is the Virasoro central charge.
Taking into account that
\begin{eqnarray}
\label{Vir L}
L_n=\oint \frac{d\xi}{2\pi i} \xi^{n+1}T(\xi)
\end{eqnarray}
we get
\begin{eqnarray}
\label{Lfff}
\langle L_{-2}\varphi_1\varphi_2\varphi_3\rangle
=\oint_{\mathit{c}(x_1)}\frac{d\xi}{2\pi i}\frac{1}{\xi-x_1}
\langle T(\xi)\varphi_1\varphi_2\varphi_3\rangle
\end{eqnarray}
and
\begin{eqnarray}
\label{LfLff}
\langle L_{-2}\varphi_1L_{-2}\varphi_2\varphi_3\rangle
=\oint_{\mathit{c}(x_1)}\frac{d\xi_1}{2\pi i}\oint_{\mathit{c}(x_2)}\frac{d\xi_2}{2\pi i}\frac{1}{\xi_1-x_1}\frac{1}{\xi_2-x_2}
\langle T(\xi_1)T(\xi_2)\varphi_1\varphi_2\varphi_3\rangle
\end{eqnarray}
where $\mathit{c}(x_{1})$, $\mathit{c}(x_{2})$ are small
contours surrounding respectively the points $x_1$ and
$x_2$ in anti-clockwise direction.
Using (\ref{My L}), (\ref{Tfff}), (\ref{TTfff}),
(\ref{Lfff}) and (\ref{LfLff}) we obtain that
\begin{eqnarray}
\label{mLfff}
&&\langle \mathcal{L}_{2}\varphi_1(x_1,\bar{x}_1)
\varphi_2(x_2,\bar{x}_2)\varphi_3(x_3,\bar{x}_3) \rangle= \\
&&-\frac{1}{3} C_{123}\left(x_1-x_2\right)
{}^{-2-\Delta _1-\Delta _2+
\Delta _3} \left(x_1-x_3\right){}^{-2-\Delta _1+
\Delta _2-\Delta _3} \left(x_2-x_3\right){}^{2+\Delta _1-
\Delta _2-\Delta _3}\nonumber\\
&&\left(\bar{x}_1-\bar{x}_2\right){}^{-\Delta _1-\Delta _2+
\Delta _3} \left(\bar{x}_1-\bar{x}_3\right){}^{-\Delta _1+
\Delta _2-\Delta _3} \left(\bar{x}_2-\bar{x}_3\right){}^{\Delta _1-
\Delta _2-\Delta _3}\nonumber\\
&&\left(-\Delta _1+
\Delta _2+\Delta _3+2 \Delta _1 \Delta _2+2
\Delta _1 \Delta _3+6 \Delta _2 \Delta _3+\Delta _1^2-3 \Delta _2^2-3 \Delta _3^2\right)\nonumber
\end{eqnarray}
and
\begin{eqnarray}
\label{mLfmLff}
&&\langle \mathcal{L}_{2}\varphi_1(x_1,\bar{x}_1)
\mathcal{L}_{2}\varphi_2(x_2,\bar{x}_2)
\varphi_3(x_3,\bar{x}_3) \rangle=\\
&&-\frac{1}{9}C_{123} \left(x_1-x_2\right)
{}^{-4-\Delta _1-\Delta _2+\Delta _3}
\left(x_1-x_3\right){}^{-\Delta _1+\Delta _2-\Delta _3}
\left(x_2-x_3\right){}^{\Delta _1
-\Delta _2-\Delta _3}\nonumber\\
&&\left(\bar{x}_1-\bar{x}_2\right){}^{-\Delta _1-\Delta _2+\Delta _3}
\left(\bar{x}_1-\bar{x}_3\right){}^{-\Delta _1+\Delta _2-\Delta _3}
\left(\bar{x}_2-\bar{x}_3\right)
{}^{\Delta _1-\Delta _2-\Delta _3}\nonumber\\
&&\left(
-2c+10 \Delta _1+10 \Delta _2+10 \Delta _3-4 c \Delta _1-4 c \Delta _2
-8c \Delta _1 \Delta _2+74 \Delta _1 \Delta _2
+44 \Delta _1 \Delta _3\right.\nonumber\\
&&+44\Delta _2 \Delta _3-33 \Delta _1^2 -33\Delta _2^2-43 \Delta _3^2
+28\Delta _1 \Delta _2 \Delta _3-40 \Delta _1 \Delta _2^2-40 \Delta _1^2 \Delta _2
+18\Delta _1^2 \Delta _3\nonumber\\
&&+18\Delta _2^2 \Delta _3-68\Delta _1 \Delta _3^2
-68\Delta _2 \Delta _3^2+8 \Delta _1^3+8 \Delta _2^3+42 \Delta _3^3
-28\Delta _1 \Delta _2 \Delta _3^2-14 \Delta _1^2\Delta _2^2\nonumber\\
&&\left.-18\Delta _1^2 \Delta _3^2-18 \Delta _2^2 \Delta _3^2
+24\Delta _1\Delta _3^3+24 \Delta _2 \Delta _3^3+4 \Delta _1 \Delta _2^3
+4 \Delta _1^3 \Delta _2+3\Delta _1^4+3\Delta _2^4-9 \Delta _3^4\right)\nonumber
\end{eqnarray}
where $C_{123}$ is the structure constant related to
the three-point function\\ $\langle\varphi_1(x_1,\bar{x}_1)
\varphi_2(x_2,\bar{x}_2)
\varphi_3(x_3,\bar{x}_3)\rangle$.
Integrals over three point functions can be calculated
using the formula
\begin{eqnarray}
\label{Tool 1}
\int d^2y\, y^{\alpha_1-1} \bar{y}^{\alpha_2-1}
(1-y)^{\beta_1-1}(1-\bar{y})^{\beta_2-1}
=\pi\frac{\Gamma(\alpha_1)\Gamma(\beta_1)
\Gamma(1-\alpha_2 -\beta_2)}{\Gamma(1-\alpha_2)\Gamma(1-\beta_2)
\Gamma(\alpha_1 +\beta_1)}
\end{eqnarray}
where the numbers  $\alpha_1-\alpha_2$ and
$\beta_1-\beta_2$ are supposed to be integers. Then we calculate the constants $\tilde{C}_{ijk}$
following the line described in section \ref{adjustment}.
\section{Calculation of the matrix of anomalous dimensions}
\label{anomdim}
First let us recall that our expression
(\ref{My gamma}) is derived with the assumption that
the fields satisfy the normalization condition
(\ref{norm}). Orthogonality of the fields (\ref{My f})
follows from the fact that they are primaries, quasi
primaries or their derivatives.
The normalization of the quasi primary fields can be
easily fixed through the replacement $\Delta_1=\Delta_2,
\Delta_3=0$ in (\ref{mLfmLff}) which reduces the three point
functions to two point ones.
For the normalization constants $N_i$ we get the
following values
\begin{eqnarray}
\label{N}
N_1 &=&N_{10}=1 \nonumber\\
N_2 &=&4 \Delta _{(n,n+1)}^2 \left(2 \Delta _{(n,n+1)}
+1\right){}^2 \left(2 \Delta _{(n,n+1)}
+2\right){}^2 \left(2 \Delta _{(n,n+1)}+3\right){}^2\nonumber\\
N_3 &=&N_4=\frac{4}{9} \Delta _{(n,n+1)}
\left(2 \Delta _{(n,n+1)}+1\right){}^2 \left(2
\Delta _{(n,n+1)}+2\right) \left(2 \Delta _{(n,n+1)}
+3\right)\nonumber\\
&\times&\left(2 c \Delta _{(n,n+1)}+16
\Delta _{(n,n+1)}^2-10 \Delta _{(n,n+1)}+c\right)\nonumber\\
N_5 &=&\frac{4}{81} \left(2 \Delta _{(n,n+1)}
+1\right){}^2 \left(2 c \Delta _{(n,n+1)}
+16 \Delta _{(n,n+1)}^2-10 \Delta _{(n,n+1)}
+c\right){}^2\nonumber\\
N_6 &=&4 \Delta _{(n,n-1)}^2 \left(2 \Delta _{(n,n-1)}
+1\right){}^2 \left(2 \Delta _{(n,n-1)}
+2\right){}^2 \left(2 \Delta _{(n,n-1)}+3\right){}^2\nonumber\\
N_7 &=&N_8=\frac{4}{9} \Delta _{(n,n-1)}
\left(2 \Delta _{(n,n-1)}+1\right)
{}^2 \left(2 \Delta _{(n,n-1)}+2\right)
\left(2 \Delta _{(n,n-1)}+3\right)\nonumber\\
&\times&\left(2 c \Delta _{(n,n-1)}+16
\Delta _{(n,n-1)}^2-10 \Delta _{(n,n-1)}+c\right)\nonumber\\
N_9 &=&\frac{4}{81} \left(2 \Delta _{(n,n-1)}
+1\right){}^2 \left(2 c \Delta _{(n,n-1)}
+16 \Delta _{(n,n-1)}^2-10 \Delta _{(n,n-1)}+c\right){}^2
\end{eqnarray}
where the central charge $c$ and the dimensions
$\Delta _{(n,m)}$ are given by (\ref{Ccharge}) and
(\ref{Kac_spectrum}).
Now we have all necessary ingredients to start the calculation
of the matrix of anomalous dimensions given by the expression
(\ref{My gamma}).
Remind that ${\cal M}_{p,p-1}$ has only one coupling
constant denoted as $g$. The perturbing field we simply denote
by $\varphi_{13}\equiv \Phi$ and its dimension as
$\Delta_{(1,3)}\equiv \Delta$. Accordingly we will suppress
the index $k$ in quantities $\tilde{C}_{ijk}$ and
$I_{ijk}$. Specializing the discussion of the section
\ref{adjustment} into our case we see that
the constants $\tilde{C}_{ij}$ should be derived
from the equality
\begin{eqnarray}
\label{My I2}
\int\mathrm{d^2}y\langle\Phi_i(x)
\Phi_j(0)\Phi(y)\rangle|_{g=0}
=(x^2)^{1-\Delta_i-\Delta_j-\Delta}
\tilde{C}_{ij}I_{ij}
\end{eqnarray}
where $I_{ij}$ is given by the eq. (\ref{Iijk}) with $\Delta_k$
replaced by $\Delta$.

As an example let us demonstrate the calculation of $\tilde{C}_{47}$.
We need the correlation function
\begin{eqnarray}
&&\langle
\mathcal{L}_2\bar{\mathcal{L}}_1
\varphi_{n,n+1}(x_1,\bar{x}_1)
\mathcal{L}_1\bar{\mathcal{L}}_2
\varphi_{n,n-1}(x_2,\bar{x}_2)
\varphi_{1,3}(y,\bar{y})\rangle\nonumber\\
&&=\frac{\partial^2}{\partial\bar{x_1}^2}
\frac{\partial^2}{\partial {x_2}^2}
\langle\mathcal{L}_2
\varphi_{n,n+1}(x_1,\bar{x}_1)
\bar{\mathcal{L}}_2
\varphi_{n,n-1}(x_2,\bar{x}_2)
\varphi_{1,3}(y,\bar{y})\rangle
\end{eqnarray}
Using the factorization property and the eq. (\ref{mLfff})
we get
\begin{eqnarray}
&&\langle\mathcal{L}_2
\varphi_{n,n+1}(x_1,\bar{x}_1)
\bar{\mathcal{L}}_2
\varphi_{n,n-1}(x_2,\bar{x}_2)
\varphi_{1,3}(y,\bar{y})\rangle\nonumber\\
&&=a_1 a_2
C_{(n,n+1)(n,n-1)(1,3)}
x_{12}^{\Delta_{(1,3)}-\Delta_{(n,n-1)}-\Delta_{(n,n+1)}-2}
\bar{x}_{12}^{\Delta_{(1,3)}-\Delta_{(n,n-1)}
-2-\Delta_{(n,n+1)}}\nonumber\\
&&\times (x_1-y)^{\Delta_{(n,n-1)}
-\Delta_{(n,n+1)}-2-\Delta_{(1,3)}}
(x_2-y)^{\Delta_{(n,n+1)}+2
-\Delta_{(n,n-1)}-\Delta_{(1,3)}}
\nonumber\\
&&\times (\bar{x_1}-\bar{y})^{\Delta_{(n,n-1)}+2
-\Delta_{(n,n+1)}-\Delta_{(1,3)}}
(\bar{x_2}-\bar{y})^{\Delta_{(n,n+1)}
-\Delta_{(n,n-1)}-2-\Delta_{(1,3)}}
\end{eqnarray}
where
\begin{eqnarray}
\label{a1a2}
&&a_{1,2}=-\frac{1}{3}\left( -\Delta_{(n,n\pm1)}
+ \Delta^{2}_{(n,n\pm1)} + \Delta_{(n,n\mp1)}
+ 2 \Delta_{(n,n\pm1)}\Delta_{(n,n\mp1)}\right.\nonumber \\
&&\qquad\qquad -\left. 3\Delta^{2}_{(n,n\mp1)} + \Delta_{(1,3)}
+2\Delta_{(n,n\pm1)}\Delta_{(1,3)}
+6\Delta_{(n,n\mp1)}\Delta_{(1,3)} -
3\Delta^{2}_{(1,3)}\right)
\end{eqnarray}
while the structure constant (see(\ref{StrConstGen}))
\begin{eqnarray}
\label{Cn}
C_{(n,n+1)(n,n-1)(1,3)} = \frac{(n^2-1)^{1/2}}{\sqrt{3}n} + O(\epsilon)
\end{eqnarray}
Performing the integration over $y$ with the help of
the eq. (\ref{Tool 1}) we finally get
\begin{eqnarray}
\label{C47}
&&\tilde{C}_{47}=
N_4^{-\frac{1}{2}}
N_7^{-\frac{1}{2}}
a_1 a_2
C_{(n,n+1)(n,n-1)(1,3)}
\nonumber \\
&&\times(-\Delta_{(1,3)}-\Delta_{(n,n-1)}-\Delta_{(n,n+1)}-1)^2
(-\Delta_{(1,3)}-\Delta_{(n,n-1)}-\Delta_{(n,n+1)}-2)^2
\nonumber\\
&&\times\frac{(-t)(-t-1)(s)(s+1)}{(t-1)(t-2)(2-s)(1-s)}
\end{eqnarray}
where
\begin{eqnarray}
s = \Delta_{(n,n+1)} - \Delta_{(n,n-1)} -\Delta_{(1,3)} +1
\nonumber \\
t = \Delta_{(n,n-1)} - \Delta_{(n,n+1)} -\Delta_{(1,3)} +1
\end{eqnarray}
It remains to insert the values of the quantities
$N_4$, $N_7$, $C_{(n,n+1)(n,n-1)(1,3)}$,
$a_1,a_2$, $s$, $t$ and the respective dimensions. So we get
\begin{eqnarray}
\label{C470f}
\tilde{C}_{47} = \frac{35 \left(n^2-4\right)^{3/2} \epsilon ^2}{144 \sqrt{3} n}+O(\epsilon^3)
\end{eqnarray}
It is obvious from (\ref{My gamma}) that only the terms of
order $O(\epsilon^0)$ in  $\tilde{C}_{ijk}$ should be kept to get the
matrix element $\gamma^i_j$ with the required accuracy.
Hence for $\gamma_{4,7}$ we put zero.

In a similar manner one can calculate all other
matrix elements. Here are the nonzero matrix elements
$\gamma_{ij}$ with index  $i\leq j$
(remind that this matrix is symmetric)
\begin{eqnarray}
\label{g}
\begin{array}{ll}
\gamma_{1,1}=\frac{9}{4}+\left(-\frac{9}{8}
-\frac{3 n}{4}\right) \epsilon +
2\pi g \frac{3 \sqrt{3} (4+n)}{2 (2+n)};&\quad
\gamma_{1,5}=2\pi g \frac{\sqrt{3} (n-1)
}{n+2}\sqrt{\frac{n+3}{n+1}};\\
\gamma_{2,2}=\frac{9}{4}+\left(-\frac{1}{8}
-\frac{n}{4}\right) \epsilon
+ 2\pi g\frac{(2+n)}{2 \sqrt{3} n};&\quad
\gamma_{2,6}= 2\pi g\frac{\sqrt{-1+n^2}}{\sqrt{3} n};\\
\gamma_{3,3}=\frac{9}{4}+\left(-\frac{1}{8}
-\frac{n}{4}\right) \epsilon
+ 2\pi g\frac{(8+n)}{2 \sqrt{3} n};&\quad
\gamma_{3,7}= 2\pi g\frac{2 \sqrt{-4+n^2}}{\sqrt{3} n};\\
\gamma_{4,4}=\frac{9}{4}+\left(-\frac{1}{8}
-\frac{n}{4}\right) \epsilon
+ 2\pi g\frac{(8+n)}{2 \sqrt{3} n};&\quad
\gamma_{4,8}= 2\pi g\frac{2 \sqrt{-4+n^2}}{\sqrt{3} n};\\
\gamma_{5,5}=\frac{9}{4}+\left(-\frac{1}{8}
-\frac{n}{4}\right) \epsilon
+ 2\pi g\frac{(8+n)^2 }{2 \sqrt{3} n (2+n)};&\quad
\gamma_{5,9}= 2\pi g\frac{4 \left(-4+n^2\right)}
{\sqrt{3} n \sqrt{-1+n^2}};\\
\gamma_{6,6}=\frac{9}{4}+\left(-\frac{1}{8}
+\frac{n}{4}\right) \epsilon
+ 2\pi g\frac{(-2+n) }{2 \sqrt{3} n};&\quad
\gamma_{7,7}=\frac{9}{4}+\left(-\frac{1}{8}
+\frac{n}{4}\right) \epsilon
+ 2\pi g\frac{(-8+n)}{2 \sqrt{3} n};\\
\gamma_{8,8}=\frac{9}{4}+\left(-\frac{1}{8}
+\frac{n}{4}\right) \epsilon
+ 2\pi g\frac{(-8+n)}{2 \sqrt{3} n};&\quad
\gamma_{9,9}=\frac{9}{4}+\left(-\frac{1}{8}
+\frac{n}{4}\right) \epsilon +
2\pi g\frac{(-8+n)^2 }{2 \sqrt{3} (-2+n) n};\\
\gamma_{9,10}= 2\pi g\frac{\sqrt{3} (n+1)
}{n-2}\sqrt{\frac{n-3}{n-1}};&\quad
\gamma_{10,10}=\frac{9}{4}+\left(-\frac{9}{8}
+\frac{3 n}{4}\right) \epsilon
+ 2\pi g\frac{3 \sqrt{3} (-4+n)}{2 (-2+n)}
\end{array}
\end{eqnarray}
At the point $g=g_*$ (see eq. (\ref{2nd point})) where the conformal invariance is recovered the eigenvalues of this matrix are
\begin{eqnarray}
\label{ei.v.}
&&{\tilde \Delta}_1=\frac{9}{4}+\left(\frac{9}{8}+
\frac{3 n}{4}\right)\epsilon\nonumber\\
&&{\tilde \Delta}_2={\tilde \Delta}_3={\tilde \Delta}_4=
{\tilde \Delta}_5=\frac{9}{4}+\left(\frac{1}{8}+\frac{n}{4}\right) \epsilon\nonumber\\
&&{\tilde \Delta}_6={\tilde \Delta}_7={\tilde \Delta}_8=
{\tilde \Delta}_9=\frac{9}{4}+\left(\frac{1}{8}
-\frac{n}{4}\right) \epsilon\nonumber\\
&&{\tilde \Delta}_{10}=\frac{9}{4}+\left(\frac{9}{8}-
\frac{3 n}{4}\right)\epsilon
\end{eqnarray}
It is not difficult to see that the first eigenvalue
corresponds to $\varphi^{(p-1)}_{(n+3,n)}$,
next four eigenvalues correspond to the descendants
$\mathcal{L}_a \bar{\mathcal{L}}_b \varphi^{(p-1)}_{(n+1,n)}$,
further come the four descendants $\mathcal{L}_a \bar{\mathcal{L}}_b \varphi^{(p-1)}_{(n-1,n)}$ and the last eigenvalue
corresponds to $\varphi^{(p-1)}_{(n-3,n)}$.
Since the first and the last eigenvalues are non-degenerated,
the corresponding eigenvectors are fixed uniquely (we assume
standard unite normalization)
\begin{eqnarray}
\label{UVIRmap}
\varphi^{(p-1)}_{(n+3,n)}(x)&=&\frac{6}{n (n+1) (n+2)}
\Phi_{1}(x,g_*)+\frac{6 \sqrt{\frac{n+3}{n+1}}}{n (n+2)}
\Phi_{5}(x,g_*)
\nonumber\\
&+&
\frac{3 \sqrt{(n+3) (n-1)}}{n (n+1)}\Phi_{9}(x,g_*)+
\frac{\sqrt{n^2-9}}{n} \Phi_{10}(x,g_*)\nonumber\\
\varphi^{(p-1)}_{(n-3,n)}(x)&=&\frac{\sqrt{n^2-9}}{n}
\Phi_{1}(x,g_*)-\frac{3 \sqrt{(n-3) (n+1)}}{(n-1)n}
\Phi_{5}(x,g_*)
\nonumber\\
&+&
\frac{6 \sqrt{\frac{n-3}{n-1}}}{(n-2)n}\Phi_{9}(x,g_*)-
\frac{6}{n(n-1)(n-2)}\Phi_{10}(x,g_*)
\end{eqnarray}
\section{Comparison with RG domain wall approach}
\label{Comparison}
Denote our $10$ properly normalized fields of the ultraviolet
theory as $\Phi_\alpha(x,0)\equiv \varphi_\alpha(x)$. Then
the one-point functions of the product theory with
Gaiotto's boundary condition are of the form
\begin{eqnarray}
\label{RG}
\langle \varphi^{(p-1)}_{n\pm 3,n} \varphi_\alpha| RG \rangle
= M^{(\pm)}_\alpha\frac{\sqrt{S^{(p-3)}_{1,n\pm 3}
S^{(p-1)}_{1,n_\alpha}}}{S^{(p-2)}_{1,n}}
\end{eqnarray}
where
\begin{eqnarray}
n_1=n+3;\quad n_2=n_3=n_4=n_5=n+1;\quad n_6=n_7=n_8=n_9=n-1;
\quad n_{10}=n-3
\end{eqnarray}
The modular matrix of the $SU(2)_k$ WZW model
is given by
\begin{eqnarray}
S^{(k)}_{m,n}=\sqrt{\frac{2}{k+2}} \sin
\left(\frac{\pi nm}{k+2}\right)
\end{eqnarray}
and the coefficients $M^{(\pm)}_\alpha$ have been computed
following the algorithm described in \cite{Gaiotto:2012RGDW}.
The computation is rather lengthy, e.g. to calculate the
coefficient $M_1$ one should consider $42$ dimensional space
of level $9/2$ fields in the theory ${\cal SM}_{k+1}
\times {\cal M}_{3}$ (${\cal SM}_{k+1}$ is the $N=1$
super-conformal series \cite{Eich:1985SCFT,
BKT:1985SCFT,FQS:1985SCFT,ZP:1988SCFT}, $k\equiv p-2$,
${\cal M}_{3}$ is the Ising model).
Here we present only the
final results, leaving details of computations for a
later publication.
\begin{eqnarray}
\begin{array}{l}
M^{(+)}_1=\left(35 \left(2 k^2+8 k+3\right) n^4+5 \left(52 k^2
+208 k+159\right) n^3\right.\\
\qquad +\left(-71 k^4-568 k^3-1299 k^2
-652 k+270\right) n^2\\
\qquad +\left(-133 k^4-1064 k^3-2907 k^2
-3116 k-1080\right) n\\
\left.\qquad +6 k (k+1) (k+3) (k+4)
(2 k+3) (2 k+5)-15 n^6-75 n^5\right)\\
\qquad \times \frac{1}{(n+1) (n+2) (n+3) (k-n) (k-n+1)
(2 k-n+3) (k+n+3) (k+n+4) (2 k+n+5)}\\
\vspace{0.1cm}\\
M^{(+)}_2=\left(\left(-6 k^2-28 k-33\right) n^5
-(k+4) \left(10 k^2+40 k+39\right) n^4\right.\\
\qquad -2 (k+3) \left(3 k^3+20 k^2+44 k+35\right) n^3
-2 \left(7 k^5+83 k^4+400 k^3+978 k^2+1206 k+588\right) n^2\\
\qquad -(k+3) \left(12 k^5+166 k^4
+863 k^3+2135 k^2+2493 k+1095\right) n\\
\qquad \left. -3 (k+1) (k+3)^2 (k+4) (2 k+3) (3 k+5)\right)\\
\qquad\times \frac{1}{(k+2)^2 (n+1) (k-n+1)
(k+n+3) (k+n+4) (2 k+n+5) \left(3 k^2-2 k
(n-7)+(n-4) n+15\right)}\\
\vspace{0.1cm}\\
M^{(+)}_3=M^{(+)}_4=\frac{\sqrt{2} \left(\frac{k+3}{k+2}\right)^{3/2}
(n-1) (k-n+3) (2 k-n+6) (3 k+n+5) }{(n+1)
\sqrt{(n-1) (n+2) (k-n+1) (k-n+3) (2 k-n+3)
(2 k-n+6)}}\\
\qquad\qquad\times\frac{\left(4 k^3+k^2(5 n+26)+k (17 n+46)-(n-3)
(n+2) (n+4)\right)}{(k+n+3) (k+n+4) (2 k+n+5)
\left(3 k^2-2 k (n-7)+(n-4) n+15\right)}\\
\vspace{0.1cm}\\
M^{(+)}_5=\left(2 (k+3)^2 (k+4) (2 k+3) (3 k+5)
\left(6 k^2+27 k+26\right)\right.\\
\qquad+(k+3) \left(96 k^5+1077 k^4+4708 k^3+9995
k^2+10220 k+3996\right) n\\
\qquad+\left(-51 k^5-665 k^4-3338 k^3-8123 k^2-9653 k-4542\right) n^2\\
\qquad+\left(26 k^3+183 k^2+410 k+312\right) n^4+
\left(-60 k^4-525 k^3-1699 k^2-2406 k-1254\right) n^3\\
\qquad\left.+2 \left(2 k^2+2 k-3\right) n^5+(-7 k-18) n^6\right)\\
\qquad\times \frac{1}{(k+2) (n+1) (n+2) (2 k-n+3)(k+n+3) (k+n+4)
(2 k+n+5) \left(3 k^2-2 k
(n-7)+(n-4) n+15\right)}\\
\vspace{0.1cm}\\
M^{(+)}_6=\frac{(k+3) \left(k \left(4 k^2-2 k-59\right)
-73\right) n+(2 k (k+4)+7) n^3
+(k (6 k (k+5)+37)+1) n^2-3 (k+3)^2 (2 k+3)
(3 k+5)}{(k+2)^2 (k+n+3) (2 k+n+5)
\left(3 k^2+2 k (n+7)+n (n+4)+15\right)}\\
\vspace{0.1cm}\\
M^{(+)}_7=M^{(+)}_8=\frac{\sqrt{2} \left(\frac{k+3}{k+2}\right)^{3/2}
\sqrt{(n-2) (n+1) (k+n+1) (k+n+3) (2 k+n+3)
(2 k+n+6)} (3 k+n+5)}{(n+1) (k+n+3) (2 k+n+5)
\left(3 k^2+2 k (n+7)+n (n+4)+15\right)}\\
\vspace{0.1cm}\\
M^{(+)}_9=\frac{18 k^4+3 k^3 (7 n+55)+2 k^2
(4 n (n+19)+273)+k (n (n (n+41)+353)+769)
+2 (n (n (n+26)+130)+195)}{(k+2) (n+1) (2 k+n+5)
\left(3 k^2+2 k (n+7)+n (n+4)+15\right)}\nonumber\\
\vspace{0.1cm}\nonumber\\
M^{(+)}_{10}=1
\end{array}\\
\end{eqnarray}
The coefficients $M^{(-)}_\alpha$ can be obtained from
$M^{(+)}_\alpha$ by replacing $n\leftrightarrow -n$.
Taking the limit $k\rightarrow \infty $ we find
\begin{eqnarray}
&&M^{(\pm)}_1=\pm\frac{1}{(n\pm1)(n\pm2)(n\pm3)}+O\left(1/k\right)\nonumber\\
&&M^{(\pm)}_5=\frac{6}{(n\pm1)(n\pm2)}+O\left(1/k^2\right)\nonumber\\
&&M^{(\pm)}_9=\mp\frac{3}{(n\mp1)}+O\left(1/k^2\right)\nonumber\\
&&M^{(\pm)}_{10}=1
\end{eqnarray}
and the remaining coefficients are of order $O(1/k)$ or higher.
Inserting these values in (\ref{RG}) we get a complete agreement with (\ref{UVIRmap}).
\section*{ Acknowledgement}
We thank Rubik Poghossian for introducing us to the subject.
\begin{appendix}
\section{Minimal models of the 2d CFT}
Here we present several formulae concerning the unitary
minimal series of 2d CFT ${\cal M}_p$.
The central charge is given by
\begin{eqnarray}
\label{Ccharge}
c=1-\frac{6}{p(p+1)}
\end{eqnarray}
The primary fields are denoted by $\varphi_{n,m}$,
$n=1,2,\ldots ,p-1$, $m=1,2,\ldots ,p$ and the corresponding
dimensions are \cite{Kac}
\begin{eqnarray}
\label{Kac_spectrum}
\Delta_{(n,m)}=\frac{(n-m)^2}{4}+\frac{n^2-1}{4p}-
\frac{m^2-1}{4(p+1)}
\end{eqnarray}
Notice also that the identification $\varphi_{n,m}\equiv
\varphi_{p-n,p+1-m}$ holds.
The structure constants of the operator algebra have
been computed in \cite{DF:1985SC}. Here we present a
little bit simpler expression borrowed from
\cite{Pog:1989SC}
\begin{eqnarray}
\label{StrConstGen}
&&C_{(n_1,m_1)(n_2,m_2)(n_3,m_3)}=\rho^{4st+2t-2s-1}\nonumber\\
&&\times \sqrt{\frac{\gamma(\rho-1) \gamma(m_1-n_1\rho^{-1})
\gamma(m_2-n_2\rho^{-1})\gamma(-m_3+n_3\rho^{-1})}
{\gamma(1-\rho^{-1}) \gamma(-n_1+m_1\rho)
\gamma(-n_2+m_2\rho)\gamma(n_3-m_3\rho)}}\nonumber\\
&&\times\prod_{i=1}^s\prod_{j=1}^t\left((i-j\rho)
(i+n_3-(j+m_3)\rho)(i-n_1-(j-m_1)\rho)
(i-n_2-(j-m_2)\rho)\right)^{-2}\nonumber\\
&&\times\prod_{i=1}^s\gamma(i\rho^{-1})\gamma(-m_3+(i+n_3)\rho^{-1})
\gamma(m_1+(i-n_1)\rho^{-1})\gamma(m_2+(i-n_2)\rho^{-1})\nonumber\\
&&\times\prod_{j=1}^t\gamma(j\rho)\gamma(-n_3+(j+m_3)\rho)
\gamma(n_1+(j-m_1)\rho)\gamma(n_2+(j-m_2)\rho)
\end{eqnarray}
where
\begin{eqnarray}
\rho=\frac{p}{p+1};
\quad \gamma(x)\equiv\frac{\Gamma(x)}{\Gamma(1-x)};\quad
s=\frac{n_1+n_2-n_3-1}{2}; \qquad t=\frac{m_1+m_2-m_3-1}{2}\nonumber
\end{eqnarray}
\end{appendix}
\bibliographystyle{JHEP}
\providecommand{\href}[2]{#2}
\begingroup\raggedright

\endgroup

\end{document}